\documentclass[journal]{IEEEtran}

\usepackage{cite}
\usepackage{url}
\usepackage{graphicx}
\usepackage{color}
\usepackage{bm}
\usepackage{cite}
\usepackage{amsmath,amssymb,amsfonts}
\usepackage{algorithmic}
\usepackage{graphicx}
\usepackage{textcomp}
\usepackage{algorithmic}
\usepackage{algorithm}
\usepackage{array}
\usepackage[caption=false,font=normalsize,labelfont=sf,textfont=sf]{subfig}
\usepackage{stfloats}
\usepackage{url}
\usepackage{verbatim}
\usepackage{enumerate}
\usepackage{tikz}
\usepackage{bm}
\usepackage{threeparttable,booktabs}
\usepackage{blkarray}
\usepackage{amsmath}
\usepackage{multirow}

\usepackage{amsmath}
\usepackage{amsthm,amsfonts}
\usepackage{pgfplots}

\usepackage{tikz}
\usepackage[customcolors]{hf-tikz}
\usepgfplotslibrary{colorbrewer}
\usetikzlibrary{%
  3d,
  arrows,%
  shapes.misc,
  shapes.arrows,%
  chains,%
  matrix,%
	fit,%
  positioning,
  scopes,%
  decorations.pathmorphing,
  shadows,%
  decorations.markings,
  shapes.geometric,
  arrows.meta,bending,
	patterns,
	intersections, 
	pgfplots.fillbetween
}

\usepackage{tikz-3dplot}
\usepackage[pdfusetitle]{hyperref}
\hypersetup{
    colorlinks,
    linkcolor={red!50!black}, 
    citecolor={blue!50!black},
    urlcolor={blue!50!black}
}

\usepackage{cellspace}
\providecommand*{\mathcalbf}[1]{\bm{\mathcal{#1}}}

\renewcommand{\vec}[1]{{\boldsymbol#1}}

\DeclareMathAccent{\ring}{\mathalpha}{operators}{"17}

\newcommand{\ie}{\textit{i.e.}\/, }
\newcommand{\eg}{\textit{e.g.}\/, }

\colorlet{dpurple}{blue!50!red}
\colorlet{dblue}{blue!50!black}
\colorlet{dgreen}{green!50!black}
\colorlet{dred}{red!50!black}
\colorlet{dyellow}{yellow!50!black}
\colorlet{dorange}{orange!50!black}
\definecolor{metal}{RGB}{218,165,32}
\definecolor{diel}{RGB}{1,165,32}
\definecolor{antenna}{RGB}{100,150,162}
\definecolor{breg}{rgb}{0.2,0.6,0.8}%
\definecolor{preg}{rgb}{0.8,0.2,0.2}%
\definecolor{reg}{RGB}{218,165,32}

\graphicspath{{figures/}{tikz/}}

\makeatother

\begin{document}
\title{Hybrid Method of Moments and Generalized Scattering Matrix: Applications to Antennas in Radomes, Reflectors, and Implantable Media}

\author{Chenbo Shi, Shichen Liang, Xin Gu, Jin Pan and Le Zuo
\thanks{Manuscript received Sep. 25, 2025. (\textit{Corresponding author: Jin Pan.})}
\thanks{Chenbo Shi, Shichen Liang, Xin Gu and Jin Pan are with the School of Electronic Science and Engineering, University of Electronic Science and Technology of China, Chengdu 611731 China  (e-mail: chenbo\_shi@163.com; lscstu001@163.com; xin\_gu04@163.com; panjin@uestc.edu.cn).}
\thanks{Le Zuo is with The 29th Research Institute of China Electronics Technology Group Corporation (e-mail: zorro1204@163.com)}
}


\maketitle

\begin{abstract}
Electromagnetic analysis of antennas embedded in or interacting with large surrounding structures poses inherent multiscale challenges: the antenna is electrically small yet geometrically detailed, while the environment is electrically large but comparatively smooth. To address this, we present a hybrid method of moments (MoM) and generalized scattering matrix (GSM) framework that achieves a clean separation between fine-scale and large-scale complexities while preserving their full mutual coupling. Antennas of arbitrary geometry can be characterized once and reused across different environments, or conversely, a given environment can be modeled once to accommodate multiple antenna designs. The framework is inherently versatile, encompassing GSM-PO and GSM + T-matrix extensions, and thus provides a unified paradigm for multiscale antenna modeling. With the large body always represented by the formulation best suited to its scale and shape, the approach combines accuracy, efficiency, and adaptability. Numerical validations on implantable antennas, radome-protected arrays, and reflector systems confirm excellent agreement with full-wave solvers while demonstrating dramatic reductions in computational cost for design and optimization.
\end{abstract}

\begin{IEEEkeywords}
  The method of moments (MoM), generalized scattering matrix (GSM), hybrid method, radomes, implantable antennas.
\end{IEEEkeywords}

\section{Introduction}

\IEEEPARstart{E}{lectromagnetic} problems involving antennas embedded in or interacting with large surrounding structures frequently exhibit strong multiscale characteristics. Typical scenarios include implantable antennas inside biological tissues, reflector antennas illuminated by compact feeds, and radome-enclosed antenna arrays. In such scenarios, the antenna itself is typically small or moderately sized but contains intricate details, while the surrounding structure is electrically large yet comparatively smooth. Developing numerical methods that can capture the fine-scale details of the antenna while efficiently accounting for the large-scale environment remains an enduring challenge in computational electromagnetics.

Hybrid methods have long been recognized as effective strategies for tackling multiscale problems \cite{ref_FEM_MoM_hyb1,ref_FEM_MoM_hyb2,ref_FEM_MoM_hyb3,ref_FEM_MoM_hyb4}, as they allow different subsystems to be modeled by the most suitable formulations. A notable example is the hybrid approach of method-of-moments (MoM) and T-matrix \cite{ref_hybrid}, which provides an elegant treatment of domain-to-domain coupling at their boundaries. In this method, the large body is represented by a spherical-wave expansion, while the fine structure is modeled by a surface-current formulation (MoM). This formulation is well suited for canonical shapes or quasi-spherical scatterers, but it becomes less effective when the surrounding structure is electrically large and irregular. In such cases, acceleration techniques such as the multilevel fast multipole method (MLFMM) \cite{ref_MLFMA} must be employed to obtain the T-matrix; yet, because the computation inevitably involves multiple right-hand sides, the achievable efficiency is essentially limited to $\mathcal{O}(N^{2}\log N)$---as analyzed in detail in Sec.~\ref{Sec_IIIC}---rendering the approach less attractive for electrically large radomes or reflectors.

In this work, we propose a new MoM and generalized scattering matrix (GSM) hybrid framework, in which the roles of antenna and environment are inverted compared with \cite{ref_hybrid}. Specifically, the electrically large environment---radome, reflector, or implantation medium---is discretized and solved by the MoM, while the antenna is described by its GSM \cite{ref_myGSM,ref_sph_near_measure}. This treatment provides two distinct benefits. First, the antenna's geometric details are encapsulated in a compact GSM representation, avoiding their direct inclusion in the MoM system and thereby reducing the number of unknowns. Second, the framework can seamlessly exploit recent advances in GSM theory, including analytic manipulations based on Wigner translation and rotation theorems \cite{ref_mysyn_CMA,ref_mysyn_GSM}, efficient array synthesis from single-element GSMs \cite{ref_GSM_3DFEM,ref_GSM_CMA,ref_GSM_nearsph}, and modular reuse of precomputed antenna data across different scenarios.

A further strength of the proposed framework lies in its flexibility to evolve into specialized variants. By reformulating the coupling operators, the same theoretical foundation can be extended into: (i) a GSM-PO hybrid for large reflectors where physical optics yields fast current estimates. Unlike the simplified formulation in \cite{ref_GSM_PO}, which neglects multiple reflections, the present GSM-PO hybrid is derived in a full-coupled form and thus captures higher-order interactions between the antenna and reflector; and (ii) a GSM + T-matrix hybrid, advantageous when the surrounding structure itself admits an efficient spherical-wave expansion. The latter inherits most of the advantages in \cite{ref_hybrid}, while further leveraging the modularity of GSM to accommodate more general antenna-environment interactions. Each of these hybrids exhibits unique advantages for its target scenario, together forming a versatile family of tools for multiscale antenna analysis.

Beyond serving as a specific computational scheme, the proposed framework establishes a unifying paradigm for multiscale electromagnetic analysis. By reconciling fine-scale antenna details with electrically large environments within a modular GSM-MoM architecture, it bridges full-wave rigor with system-level efficiency. This perspective highlights not only the accuracy and efficiency of the method, but also its universality: antennas and environments can be modeled and reused in a plug-and-play fashion, and different large-scale structures can be represented by the formulation most natural to their geometry and scale. As such, the framework lays a foundation for the practical design and optimization of advanced antenna systems---from implantable devices to radome-protected arrays and reflector antennas---that would be prohibitively expensive for conventional full-wave approaches.

\section{MoM and GSM Hybrid Formula}
\label{SecII}

\begin{figure}[!t]
  \centering
  \includegraphics[]{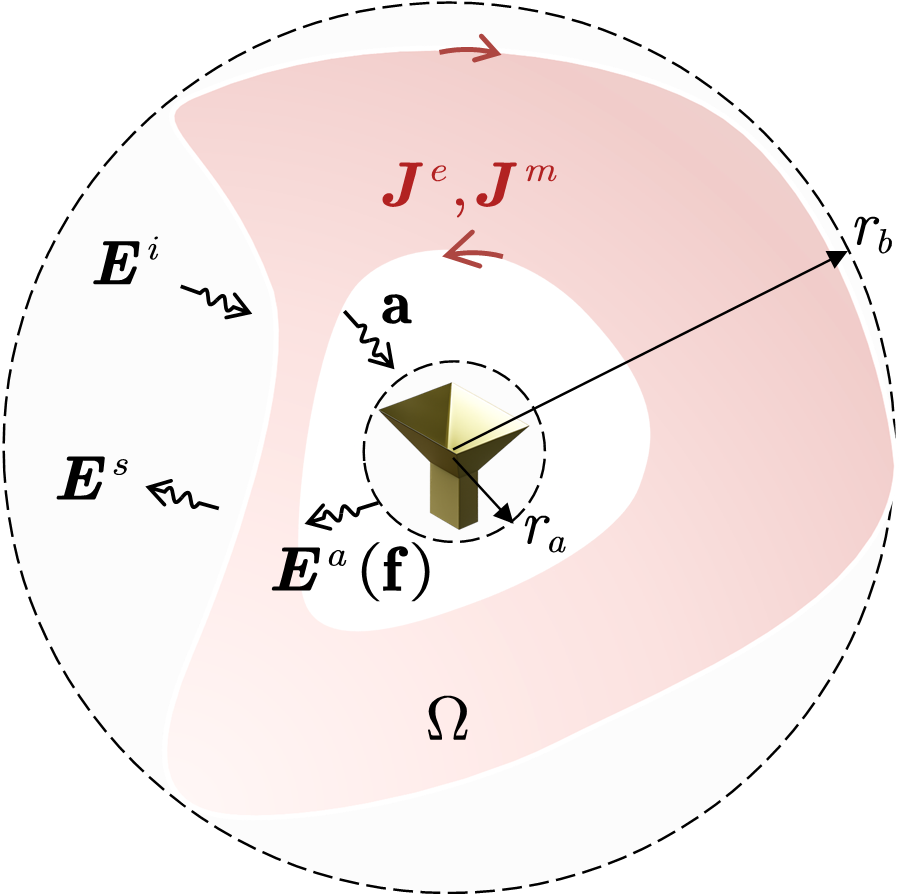}
  \caption{Decomposition of the total electric field when the antenna is embedded inside a penetrable medium. The total field in free space (both inside and outside) is expressed as the superposition of $\vec E^i$, $\vec E^s$ and $\vec E^a$. The vector $\mathbf{f}$ denotes the spherical-wave expansion coefficients of $\vec E^a$, while $\mathbf{a}$ denotes those of $\vec E^i+\vec E^s$; they serve as notational representations rather than additional physical fields. The radii $r_a$ and $r_b$ correspond to the smallest spheres enclosing the antenna and the structure $\Omega$, respectively.}
  \label{f_EMmdel_int}
\end{figure}

As illustrated in Fig.~\ref{f_EMmdel_int}, the free space is partitioned by a penetrable structure $\Omega$ into interior and exterior regions. An incident electromagnetic wave, $\boldsymbol{E}^i, \boldsymbol{H}^i$ (\eg a uniform plane wave), impinges upon the system. An antenna is placed in the interior region, acting both as a scatterer and as a radiator. The total fields in free space (covering both the interior and exterior regions) can thus be expressed as
\begin{equation}
  \boldsymbol{E}=\boldsymbol{E}^i+\boldsymbol{E}^s+\boldsymbol{E}^a,\boldsymbol{H}=\boldsymbol{H}^i+\boldsymbol{H}^s+\boldsymbol{H}^a.
\end{equation}
Here, $\boldsymbol{E}^s, \boldsymbol{H}^s$ denote the fields generated by the structure $\Omega$, while $\boldsymbol{E}^a, \boldsymbol{H}^a$ are produced by the antenna. The scattered fields $\boldsymbol{E}^s, \boldsymbol{H}^s$ originate from the radiation of equivalent electric and magnetic currents, $\boldsymbol{J}^e, \boldsymbol{J}^m$, induced on the penetrable structure $\Omega$ under the combined excitation of $\boldsymbol{E}^i, \boldsymbol{H}^i$ and $\boldsymbol{E}^a, \boldsymbol{H}^a$. These scattered fields can be expressed as
\begin{equation}
  \label{eq2}
  \begin{split}
   &\boldsymbol{E}^s=-\mathrm{j}k_0Z_0\mathcal{L} _0\left( \boldsymbol{J}^e \right) -\mathcal{K} _0\left( \boldsymbol{J}^m \right) \\
   &\boldsymbol{H}^s=\mathcal{K} _0\left( \boldsymbol{J}^e \right) -\frac{\mathrm{j}k_0}{Z_0}\mathcal{L} _0\left( \boldsymbol{J}^m \right) 
  \end{split}
\end{equation}
where $k_0$ and $Z_0$ denote the free-space wavenumber and impedance, respectively. The operators $\mathcal{L}, \mathcal{K}$ are defined as
\begin{equation}
  \begin{split}
    &\mathcal{L} _0\left\{ \boldsymbol{J} \right\} \left( \boldsymbol{r} \right) =\left< \bar{\mathbf{G}}_0\left( \boldsymbol{r},\boldsymbol{r}^{\prime} \right) ,\boldsymbol{X}\left( \boldsymbol{r}^{\prime} \right) \right> _{\Omega ^{\prime}}    \\
    &\mathcal{K} _0\left\{ \boldsymbol{J} \right\} \left( \boldsymbol{r} \right) =\nabla \times \mathcal{L} _0\left\{ \boldsymbol{J} \right\} \left( \boldsymbol{r} \right)
  \end{split}
\end{equation}
with $\bar{\mathbf{G}}_0(\boldsymbol{r},\boldsymbol{r}')$ being the free-space dyadic Green's function \cite{ref_EMtheory}. The symmetric inner product is defined as
\begin{equation}
  \left< \boldsymbol{X}\left( \boldsymbol{r} \right) ,\boldsymbol{Y}\left( \boldsymbol{r} \right) \right> _V=\int_V{\boldsymbol{X}\left( \boldsymbol{r} \right) \cdot \boldsymbol{Y}\left( \boldsymbol{r} \right) \mathrm{d}V}.
\end{equation}

To determine $\boldsymbol{J}^e, \boldsymbol{J}^m$, the Poggio-Miller-Chang-Harrington-Wu-Tsai (PMCHWT) integral equations for the penetrable medium is formulated and solved within the standard MoM framework \cite{ref_PMCHW1,ref_PMCHW2,ref_PMCHW3}. The equivalent currents are expanded using Rao-Wilton-Glisson (RWG) basis functions ${\boldsymbol{\psi}_n}$ as
\begin{equation}
  \boldsymbol{J}^e=\sum_n{I_{n}^{e}\boldsymbol{\psi }_n},\boldsymbol{J}^m=\sum_n{I_{n}^{m}\boldsymbol{\psi }_n}.
\end{equation}
Under Galerkin testing, the PMCHWT equation is cast into the matrix form
\begin{equation}
  \label{eq6}
  \mathbf{ZI}=\mathbf{V}
\end{equation}
where
\begin{equation}
  \mathbf{Z}=\begin{bmatrix}	\mathrm{j}\left( k_0Z_0\mathbf{L}_0+k_dZ_d\mathbf{L}_d \right)&		-\mathrm{j}\left( \mathbf{K}_0+\mathbf{K}_d \right)\\	-\mathrm{j}\left( \mathbf{K}_0+\mathbf{K}_d \right)&		\mathrm{j}\left( \frac{k_0}{Z_0}\mathbf{L}_0+\frac{k_d}{Z_d}\mathbf{L}_d \right)\\\end{bmatrix} 
\end{equation}
with $\mathbf{L}, \mathbf{K}$ denoting the discretized forms of the operators $\mathcal{L}, \mathcal{K}$, and the subscript $d$ indicating substitution of free-space parameters by those of the dielectric medium. The current and excitation vector are written as
\begin{equation}
  \mathbf{I}= \begin{bmatrix}	\mathbf{I}^e\\	\mathrm{j}\mathbf{I}^m\\\end{bmatrix} ,\mathbf{V}= \begin{bmatrix}	\mathbf{V}^e\\	\mathrm{j}\mathbf{V}^m\\\end{bmatrix}
\end{equation}
with
\begin{equation}
  \label{eq9}
  \mathbf{V}=\mathbf{V}^i+\mathbf{V}^a
\end{equation}
where
\begin{equation}
  \label{eq10}
  \begin{split}
      \mathbf{V}^i=\begin{bmatrix}
	    \left< \boldsymbol{\psi }_n\left( \boldsymbol{r} \right) ,\boldsymbol{E}^i\left( \boldsymbol{r} \right) \right> _{\Omega}\\
	    \mathrm{j}\left< \boldsymbol{\psi }_n\left( \boldsymbol{r} \right) ,\boldsymbol{H}^i\left( \boldsymbol{r} \right) \right> _{\Omega}\\
    \end{bmatrix},\\
    \mathbf{V}^a= \begin{bmatrix}
	  \left< \boldsymbol{\psi }_n\left( \boldsymbol{r} \right) ,\boldsymbol{E}^a\left( \boldsymbol{r} \right) \right> _{\Omega}\\
	  \mathrm{j}\left< \boldsymbol{\psi }_n\left( \boldsymbol{r} \right) ,\boldsymbol{H}^a\left( \boldsymbol{r} \right) \right> _{\Omega}\\
    \end{bmatrix}
  \end{split}
\end{equation}
Here, the first contribution $\mathbf{V}^i$ arises from the incident fields $\boldsymbol{E}^i, \boldsymbol{H}^i$, while the second contribution $\mathbf{V}^a$ originates from the antenna-induced fields $\boldsymbol{E}^a, \boldsymbol{H}^a$.

In the free space region with $r>r_a$, the antenna fields $\boldsymbol{E}^a, \boldsymbol{H}^a$ can be expressed as a superposition of vector spherical waves \cite{ref_hybrid}, \cite[Ch. 7]{ref_Sca1}:
\begin{equation}
  \label{eq11}
  \begin{split}
    &\boldsymbol{E}^a\left( \boldsymbol{r} \right) =k_0\sqrt{Z_0}\sum_{\alpha}{f_{\alpha}\boldsymbol{u}_{\alpha}^{\left( 4 \right)}\left( k_0\boldsymbol{r} \right)}\\
    &\boldsymbol{H}^a\left( \boldsymbol{r} \right) =\mathrm{j}k_0/\sqrt{Z_0}\sum_{\alpha}{f_{\alpha}\boldsymbol{u}_{\bar{\alpha}}^{\left( 4 \right)}\left( k_0\boldsymbol{r} \right)}
  \end{split}
\end{equation}
Here, $\vec{u}_\alpha^{\left(p\right)}$ are the spherical vector waves. The superindex $p = 1$ denotes the Bessel functions of the first kind (regular waves, later associated with fields $\vec{E}^s$, $\vec{H}^s$), while $p = 4$ denotes the Hankel functions of the second kind (outgoing waves). A bar over the index $\alpha$ denotes interchanging the roles of the TE and TM waves, \ie $\boldsymbol{u}_{\bar{\alpha}}^{(p)} = k\nabla \times \boldsymbol{u}_{\alpha}^{(p)}$ \cite[Ch. 7.2]{ref_Sca1}. The maximum degree of the required spherical wave functions is determined as \cite{ref_Sph_deg_trunction}
\begin{equation}
  \label{eq12}
  L_{\max}=\lceil kr_a+\iota \sqrt[3]{kr_a}+3 \rceil 
\end{equation}
where the parameter $\iota$ controls the accuracy. $r_a$ denotes the radius of the smallest sphere enclosing the antenna structure, as shown in Fig.~\ref{f_EMmdel_int}. This results in a total of $2L_{\max}(L_{\max}+2)$ vector spherical waves.

Substituting \eqref{eq11} into \eqref{eq10} yields
\begin{equation}
  \label{eq13}
  \mathbf{V}^a=\mathbf{U}_{4}^{t}\mathbf{f}
\end{equation}
where the operator $\mathbf{U}_p$ is defined as
\begin{equation}
  \mathbf{U}_p=\left[ \mathbf{U}^{\left( p \right)},-\bar{\mathbf{U}}^{\left( p \right)} \right] 
\end{equation}
with
\begin{equation}
  \begin{split}
      &\mathbf{U}^{\left( p \right)}=k_0\sqrt{Z_0}\left[ \left< \boldsymbol{u}_{\alpha}^{\left( p \right)}\left( k_0\boldsymbol{r} \right) ,\boldsymbol{\psi }_n\left( \boldsymbol{r} \right) \right> _{\Omega} \right] \\
      &\bar{\mathbf{U}}^{\left( p \right)}=k_0/\sqrt{Z_0}\left[ \left< \boldsymbol{u}_{\bar{\alpha}}^{\left( p \right)}\left( k_0\boldsymbol{r} \right) ,\boldsymbol{\psi }_n\left( \boldsymbol{r} \right) \right> _{\Omega} \right].
  \end{split}
\end{equation}
The indices $\alpha$ and $n$ run along rows and columns, respectively.

In the antenna region, due to the sources $\vec J^e$ and $\vec J^m$ are external, the scattered fields $\boldsymbol{E}^s, \boldsymbol{H}^s$ can be expanded in terms of regular waves:
\begin{equation}
  \begin{split}
    &\boldsymbol{E}^s\left( \boldsymbol{r} \right) =k_0\sqrt{Z_0}\sum_{\alpha}{a_{\alpha}^{s}\boldsymbol{u}_{\alpha}^{\left( 1 \right)}\left( k_0\boldsymbol{r} \right)}\\
    &\boldsymbol{H}^s\left( \boldsymbol{r} \right) =\mathrm{j}k_0/\sqrt{Z_0}\sum_{\alpha}{a_{\alpha}^{s}\boldsymbol{u}_{\bar{\alpha}}^{\left( 1 \right)}\left( k_0\boldsymbol{r} \right)}.
  \end{split}
\end{equation}

Because the dyadic Green's function in the case $r<r'$ admits the expansion
\begin{equation}
  \bar{\mathbf{G}}_0\left( \boldsymbol{r},\boldsymbol{r}^{\prime} \right) =-\mathrm{j}k_0\sum_{\alpha}{\boldsymbol{u}_{\alpha}^{\left( 1 \right)}\left( k_0\boldsymbol{r} \right) \boldsymbol{u}_{\alpha}^{\left( 4 \right)}\left( k_0\boldsymbol{r}^{\prime} \right)}
\end{equation}
together with \eqref{eq2}, we obtain
\begin{equation}
  \label{eq18}
  \boldsymbol{E}^s =-k_{0}^{2}\sum_{\alpha}{\left[ Z_0\left< \boldsymbol{u}_{\alpha}^{\left( 4 \right)},\boldsymbol{J}^e \right> -\left< \boldsymbol{u}_{\bar{\alpha}}^{\left( 4 \right)},\mathrm{j}\boldsymbol{J}^m \right> \right] \boldsymbol{u}_{\alpha}^{\left( 1 \right)}}.
\end{equation}
Thus, the expansion vector satisfies
\begin{equation}
  \mathbf{a}^s=-\mathbf{U}_4\mathbf{I}.
\end{equation}

The total expansion vector of the antenna incidence waves, $\mathbf{a}$, consists of the contributions from both the scattered fields $\boldsymbol{E}^s, \boldsymbol{H}^s$ generated by the structure $\Omega$ and the incident fields $\boldsymbol{E}^i, \boldsymbol{H}^i$ in free space, \ie
\begin{equation}
  \label{eq20}
  \mathbf{a}=\mathbf{a}^s+\mathbf{a}^i=-\mathbf{U}_4\mathbf{I}+\mathbf{a}^i.
\end{equation}
For certain canonical excitations, such as uniform plane waves, the expansion coefficients $\mathbf{a}^i$ can be obtained analytically \cite[Appendix A]{ref_myGSM}.

\begin{figure}[!t]
  \centering
  \includegraphics[]{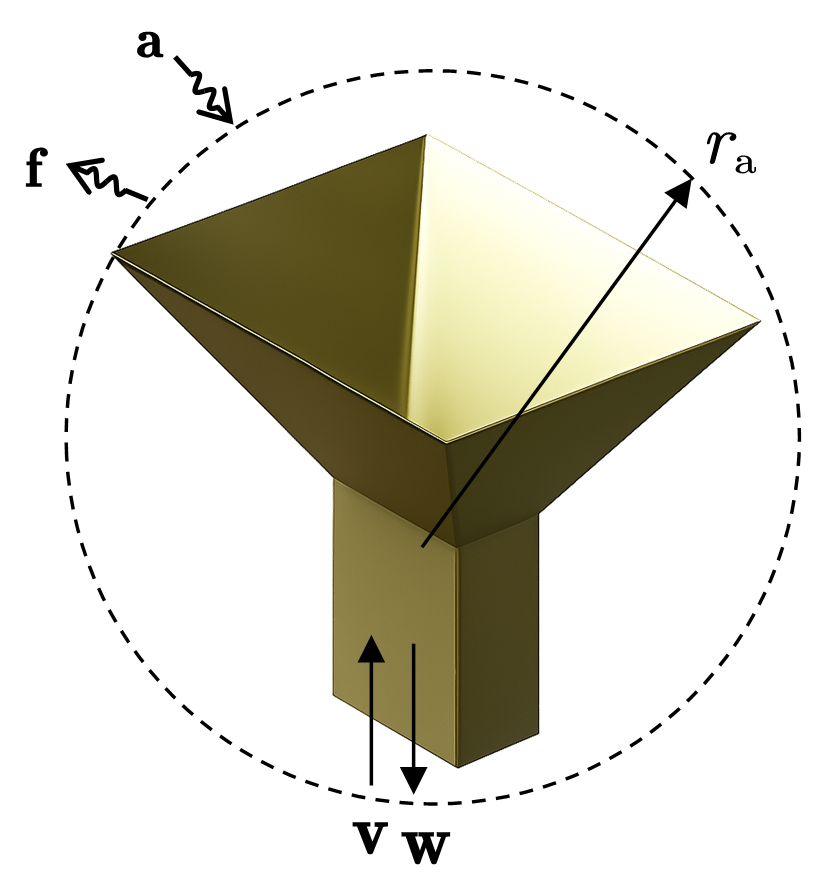}
  \caption{Illustration of antenna excitation and response. The vectors $\mathbf{a}$ and $\mathbf{v}$ represent the excitations from free space and the feeding waveguide, respectively, while $\mathbf{f}$ and $\mathbf{w}$ denote the corresponding antenna responses in free space and the waveguide. Their relations are governed by the generalized scattering matrix of the antenna.}
  \label{f_GSM_ant}
\end{figure}

The algebraic relationship between $\mathbf{a}$ and $\mathbf{f}$ can be established through the generalized scattering matrix of the antenna, formulated via the source-scattering representation \cite{ref_myGSM}, \cite[Ch. 2.3.5]{ref_sph_near_measure}. As illustrated in Fig.~\ref{f_GSM_ant}, and considering the modal expansion vectors $\mathbf{v}, \mathbf{w}$ corresponding to the guided eigenmodes of the antenna waveguide, the linear mapping is expressed as
\begin{equation}
  \label{eq21}
  \begin{bmatrix}
	\mathbf{\Gamma }&		\frac{1}{2}\mathbf{R}\\
	\mathbf{T}&		\frac{1}{2}\left( \mathbf{S}-\mathbf{1} \right)\\
  \end{bmatrix}  \begin{bmatrix}
	\mathbf{v}\\
	\mathbf{a}\\
  \end{bmatrix} = \begin{bmatrix}
	\mathbf{w}\\
	\mathbf{f}\\
  \end{bmatrix}
\end{equation}
where $\mathbf{\Gamma}, \mathbf{R}, \mathbf{T}, \mathbf{S}$ represent the port S-parameters, the receiving matrix, the transmitting matrix, and the scattering matrix, respectively. Substituting \eqref{eq20} into the second row of \eqref{eq21} gives
\begin{equation}
  \label{eq22}
  \mathbf{f}=\mathbf{Tv}+\frac{1}{2}\left( \mathbf{S}-\mathbf{1} \right) \left( -\mathbf{U}_4\mathbf{I}+\mathbf{a}^i \right).
\end{equation}
By further substituting \eqref{eq13}, \eqref{eq22}, and \eqref{eq9} into \eqref{eq6}, we obtain
\begin{equation}
  \mathbf{ZI}=\mathbf{V}^i+\mathbf{U}_{4}^{t}\left[ \mathbf{Tv}+\frac{1}{2}\left( \mathbf{S}-\mathbf{1} \right) \left( -\mathbf{U}_4\mathbf{I}+\mathbf{a}^i \right) \right]
\end{equation}
which can be reorganized as
\begin{equation}
  \label{eq24}
  \tilde{\mathbf{Z}}\mathbf{I}=\mathbf{V}^i+\mathbf{U}_{4}^{t}\frac{1}{2}\left( \mathbf{S}-\mathbf{1} \right) \mathbf{a}^i+\mathbf{U}_{4}^{t}\mathbf{Tv}
\end{equation}
with
\begin{equation}
  \label{eq25}
  \tilde{\mathbf{Z}}=\mathbf{Z}+\mathbf{U}_{4}^{t}\frac{1}{2}\left( \mathbf{S}-\mathbf{1} \right) \mathbf{U}_4.
\end{equation}

For a specific problem, $\mathbf{V}^i$, $\mathbf{a}^i$, or $\mathbf{v}$ are known. Equation \eqref{eq24} can thus be employed to solve for the equivalent currents $\mathbf{I}$ on the $\Omega$, after which \eqref{eq22} determines the expansion vector $\mathbf{f}$.

The expansion coefficients $\mathbf{w}$ of the outward-propagating eigenmodes in the antenna waveguide are obtained from the first row of \eqref{eq21}, together with \eqref{eq20}, as
\begin{equation}
  \label{eq26}
  \mathbf{\Gamma v}+\frac{1}{2}\mathbf{R}\left( -\mathbf{U}_4\mathbf{I}+\mathbf{a}^i \right) =\mathbf{w}.
\end{equation}
For the case of antenna-source excitation only, the terms $\mathbf{a}^i$ and $\mathbf{V}^i$ vanish in \eqref{eq24}, leading to
\begin{equation}
  \mathbf{I}=\tilde{\mathbf{Z}}^{-1}\mathbf{U}_{4}^{t}\mathbf{Tv}
\end{equation}
which, substituted into \eqref{eq26}, yields
\begin{equation}
  \left( \mathbf{\Gamma }-\frac{1}{2}\mathbf{RU}_4\tilde{\mathbf{Z}}^{-1}\mathbf{U}_{4}^{t}\mathbf{T} \right) \mathbf{v}=\mathbf{w}.
\end{equation}
Since this holds for any excitation vector $\mathbf{v}$, the effective S-parameters of the antenna, including the coupling with the structure $\Omega$, are then given by
\begin{equation}
  \tilde{\mathbf{\Gamma}}=\mathbf{\Gamma }-\frac{1}{2}\mathbf{RU}_4\tilde{\mathbf{Z}}^{-1}\mathbf{U}_{4}^{t}\mathbf{T}
\end{equation}

The far-field can be obtained by superimposing the scattered field $\boldsymbol{E}^s$ of the structure $\Omega$, determined by \eqref{eq2}, with the antenna radiation $\boldsymbol{E}^a$, determined by \eqref{eq11}.

It is worth noting that, although the derivation in this section has been demonstrated for a penetrable medium $\Omega$, the formulation can be readily extended to other material types. For instance, in the case of a conducting structure, it suffices to generate the impedance matrix $\mathbf{Z}$ using the EFIE \cite{ref_EFIE}, while discarding all matrix blocks associated with magnetic currents and magnetic fields.

\section{Solution and Extension of the Hybrid Method}
\label{Sec_III}

\subsection{Solution Strategy}
\label{Sec_IIIA}

A direct way to solving the algebraic system \eqref{eq24} is to compute the LU decomposition of $\tilde{\mathbf{Z}}$. However, when the structure $\Omega$ remains fixed and only the antenna is modified, the change in \eqref{eq25} is confined to the coupling term through the matrix $\mathbf{S}$, yet a full LU decomposition of the updated $\tilde{\mathbf{Z}}$ must still be carried out. For structures $\Omega$ with relatively smooth surfaces and limited geometric details, the number of RWG basis functions can be approximately estimated as proportional to $\left(k_0 r_b\right)^2$. Consequently, the computational complexity of the LU decomposition of $\tilde{\mathbf{Z}}$ scales as $\left(k_0 r_b\right)^6$, where $r_b$ denotes the radius of the smallest sphere enclosing $\Omega$, as in Fig.~\ref{f_EMmdel_int}. This makes rapid evaluation of multiple antenna--implantation configurations computationally prohibitive.

By applying the Sherman--Morrison--Woodbury formula \cite{ref_Sherman_Morrison,ref_Woodbury}, the inverse of $\tilde{\mathbf{Z}}$ can be written as
\begin{equation}
  \label{eq30}
  \tilde{\mathbf{Z}}^{-1}=\mathbf{Z}^{-1}-\mathbf{Z}^{-1}\mathbf{U}_{4}^{t}\frac{1}{2}\left( \mathbf{S}-\mathbf{1} \right) \mathbf{M}^{-1}\mathbf{U}_4
\end{equation}
where $\mathbf{M} = \mathbf{1} + \mathbf{U}_4\mathbf{Z}^{-1}\mathbf{U}_4^t \tfrac{1}{2}\big(\mathbf{S}-\mathbf{1}\big)$. If the LU decomposition of $\mathbf{Z}$ and the product $\mathbf{U}_4\mathbf{Z}^{-1}\mathbf{U}_4^t$ are precomputed and stored, only the inverse of $\mathbf{M}$ needs to be recomputed each time the antenna is modified. Its computational complexity scales as $\left(k_0 r_a\right)^6$, which depends solely on the size of the antenna structure.

Defining the scale ratio between $\Omega$ and the antenna as
\begin{equation}
  \kappa =\frac{r_b}{r_a}
\end{equation}
we deduce that, with precomputed $\mathbf{Z}^{-1}$, the complexity of evaluating \eqref{eq30} for a new configuration is reduced by a factor of $\kappa^{-6}$. In general, when $\kappa > 1.47$, at least one order of magnitude in computational cost is saved; when $\kappa > 2.16$, the savings reach two orders of magnitude (and potentially more in practice, since certain constant factors have been neglected). Given that the scenarios of interest in this work involve implantable antennas, reflector antennas, and radomes, the ratio $\kappa$ is typically large, and thus the proposed method offers substantial efficiency gains for studying multiple antenna configurations within the same structure $\Omega$.

For electrically large $\Omega$, matrix-free MoM formulations such as the MLFMM can reduce the complexity of a matrix--vector product (MVP), $\mathbf{Z}\cdot\mathbf{x}$, to $\left(k_0 r_b\right)^2 \log\left(k_0 r_b\right)$ complexity. Meanwhile, the complexity of the MVP $\mathbf{U}_4^t\tfrac{1}{2}\big(\mathbf{S}-\mathbf{1}\big)\mathbf{U}_4\cdot\mathbf{x}$ is approximately $\left(k_0 r_b\right)^2$. Therefore, the overall complexity of the MVP $\tilde{\mathbf{Z}}\cdot\mathbf{x}$ remains on the order of $\left(k_0 r_b\right)^2 \log\left(k_0 r_b\right)$. When solving \eqref{eq24} iteratively using conjugate gradient or similar methods \cite{ref_It_CG,ref_It_algorithm}, the total complexity scales as $N_{\mathrm{it}}\left(k_0 r_b\right)^2 \log\left(k_0 r_b\right)$, where $N_{\mathrm{it}}$ is the iteration count.

Although the theoretical complexity of the matrix-free acceleration matches that of solving the full problem directly with MLFMM, the proposed hybrid framework may exhibit superior conditioning for certain classes of problems. For example, when the antenna structure contains significant local geometric detail, MLFMM may not effectively accelerate the antenna portion, and convergence of the iterative solver may even fail. By contrast, the hybrid method restricts the unknowns to the structure $\Omega$, thereby avoiding potential convergence difficulties induced by the antenna discretization. Moreover, the hybrid framework paves the way for incorporating advanced GSM techniques. For instance, antenna orientation adjustments can be efficiently computed using the Wigner translation--rotation theorem \cite{ref_sph_addition1,ref_sph_addition2,ref_Kristensson_booklet,ref_rotation_realharmonics}, eliminating the need for re-simulation. Similarly, the GSM of antenna arrays may be constructed analytically from that of individual elements \cite{ref_GSM_3DFEM,ref_GSM_CMA,ref_GSM_nearsph}.

\subsection{Consistency of the External Problem and Extension to a Full-Coupled PO Algorithm}
\label{Sec_IIIB}

\begin{figure}[!t]
  \centering
  \includegraphics[]{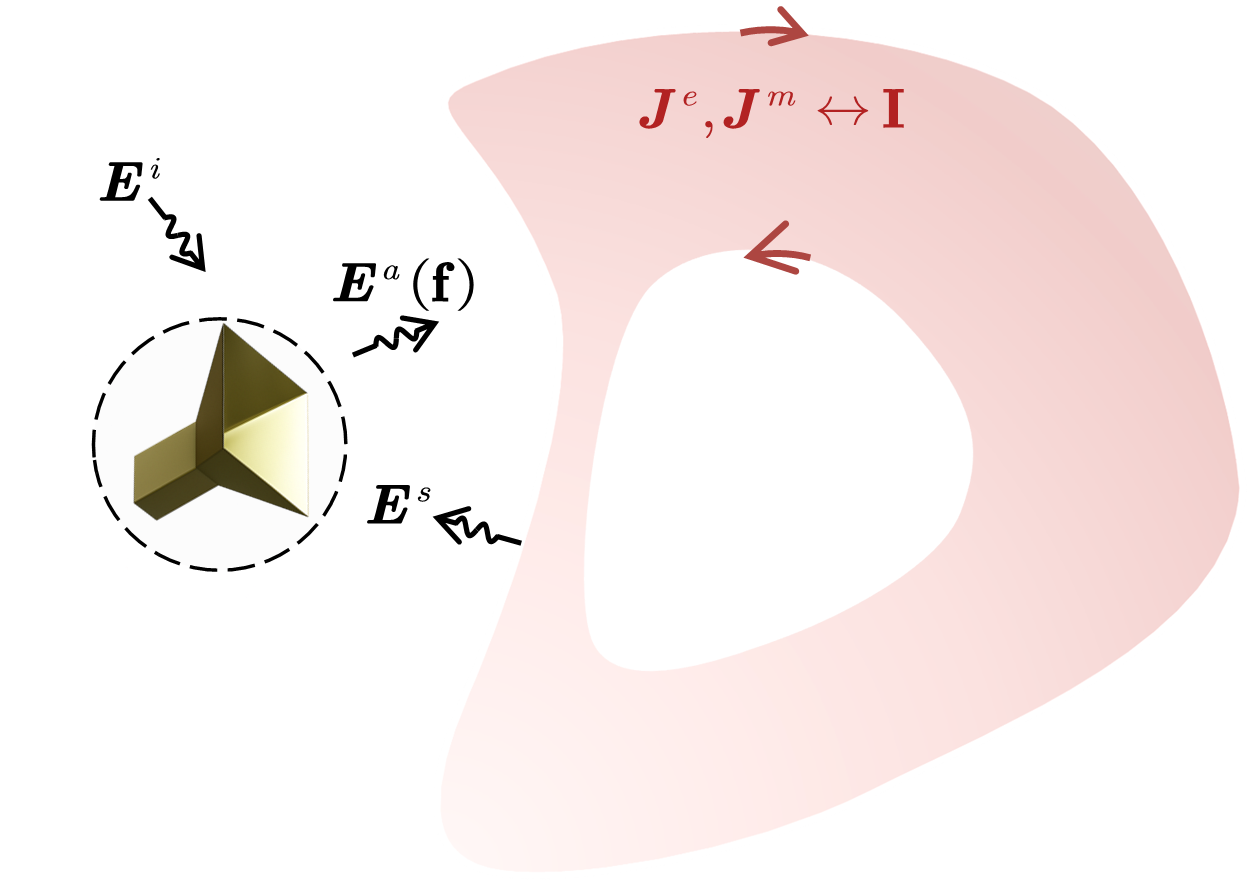}
  \caption{Field decomposition when the antenna is located outside $\Omega$. Compared with Fig.~\ref{f_EMmdel_int}, the formulation remains unchanged, except that the coordinate origin is now placed outside $\Omega$.}
  \label{f_EMmdel_ext}
\end{figure}

All the preceding derivations assumed that the antenna is placed inside the region shown in Fig.~\ref{f_EMmdel_int}. However, when the antenna is located externally, as illustrated in Fig.~\ref{f_EMmdel_ext}, the formulation remains automatically valid without modification. It should be emphasized that all matrices associated with spherical wave functions, such as $\mathbf{U}_p$, $\mathbf{a}$, and $\mathbf{f}$, are constructed in the local coordinate system of the antenna.

The configuration in Fig.~\ref{f_EMmdel_ext} demonstrates that the proposed hybrid method still covers most reflector antenna applications. A representative example is the case where $\Omega$ is a large PEC structure with small curvature, for which the physical optics (PO) approximation yields the surface current distribution as (assuming $\vec E^i,\vec H^i$ are zero)
\begin{equation}
  \boldsymbol{J}^e\left( \boldsymbol{r} \right) =2\hat{\boldsymbol{n}}\times \boldsymbol{H}^a\\=2\hat{\boldsymbol{n}}\times \frac{\mathrm{j}k_0}{\sqrt{Z_0}}\sum_{\beta}{f_{\beta}\boldsymbol{u}_{\bar{\beta}}^{\left( 4 \right)}\left( k_0\boldsymbol{r} \right)}
\end{equation}
where $\hat{\boldsymbol{n}}$ denotes the outward normal of $\Omega$. From \eqref{eq18}, the expansion coefficients of the scattered fields $\boldsymbol{E}^s, \boldsymbol{H}^s$ are obtained as
\begin{equation}
  \mathbf{a}^s=\boldsymbol{\rho }\mathbf{f}
\end{equation}
with
\begin{equation}
  \rho _{\alpha \beta}=-2\mathrm{j}k_{0}^{2}\left< \boldsymbol{u}_{\alpha}^{\left( 4 \right)},\hat{\boldsymbol{n}}\times \boldsymbol{u}_{\bar{\beta}}^{\left( 4 \right)} \right> _{\Omega}.
\end{equation}

Since $\mathbf{a}=\mathbf{a}^s=\boldsymbol{\rho}\mathbf{f}$, \eqref{eq22} reduces to
\begin{equation}
  \mathbf{f}=\mathbf{Tv}+\frac{1}{2}\left( \mathbf{S}-\mathbf{1} \right) \boldsymbol{\rho }\mathbf{f}
\end{equation}
which leads to
\begin{equation}
  \label{eq36}
  \mathbf{f}=\left[ \mathbf{1}-\frac{1}{2}\left( \mathbf{S}-\mathbf{1} \right) \boldsymbol{\rho } \right] ^{-1}\mathbf{Tv}.
\end{equation}

Substituting \eqref{eq36} into the first row of \eqref{eq21} and further using $\mathbf{a}=\boldsymbol{\rho}\mathbf{f}$, we obtain
\begin{equation}
  \left\{ \mathbf{\Gamma }+\frac{1}{2}\mathbf{R}\boldsymbol{\rho }\left[ \mathbf{1}-\frac{1}{2}\left( \mathbf{S}-\mathbf{1} \right) \boldsymbol{\rho } \right] ^{-1}\mathbf{T} \right\} \mathbf{v}=\mathbf{w}
\end{equation}
implying that the antenna reflection coefficient is given by
\begin{equation}
  \label{eq38}
  \tilde{\mathbf{\Gamma}}=\mathbf{\Gamma }+\frac{1}{2}\mathbf{R}\boldsymbol{\rho }\left[ \mathbf{1}-\frac{1}{2}\left( \mathbf{S}-\mathbf{1} \right) \boldsymbol{\rho } \right] ^{-1}\mathbf{T}.
\end{equation}

Equation \eqref{eq38} represents a full-coupled GSM-PO hybrid formulation that accounts for the multiple interactions between the antenna and the reflector. If only the first term of the Neumann series expansion of $\big[\mathbf{1}-\tfrac{1}{2}(\mathbf{S}-\mathbf{1})\boldsymbol{\rho}\big]^{-1}$ is retained, \eqref{eq38} degenerates to
\begin{equation}
  \tilde{\mathbf{\Gamma}}=\mathbf{\Gamma }+\frac{1}{2}\mathbf{R}\boldsymbol{\rho }\mathbf{T}
\end{equation}
which corresponds precisely to the conventional GSM-PO hybrid method derived in \cite{ref_GSM_PO}. Hence, \cite{ref_GSM_PO} appears as a special case of our more general full-coupled framework.

\subsection{Extension to a GSM + T-Matrix Hybrid Scheme}
\label{Sec_IIIC}

\begin{figure}[!t]
  \centering
  \includegraphics[]{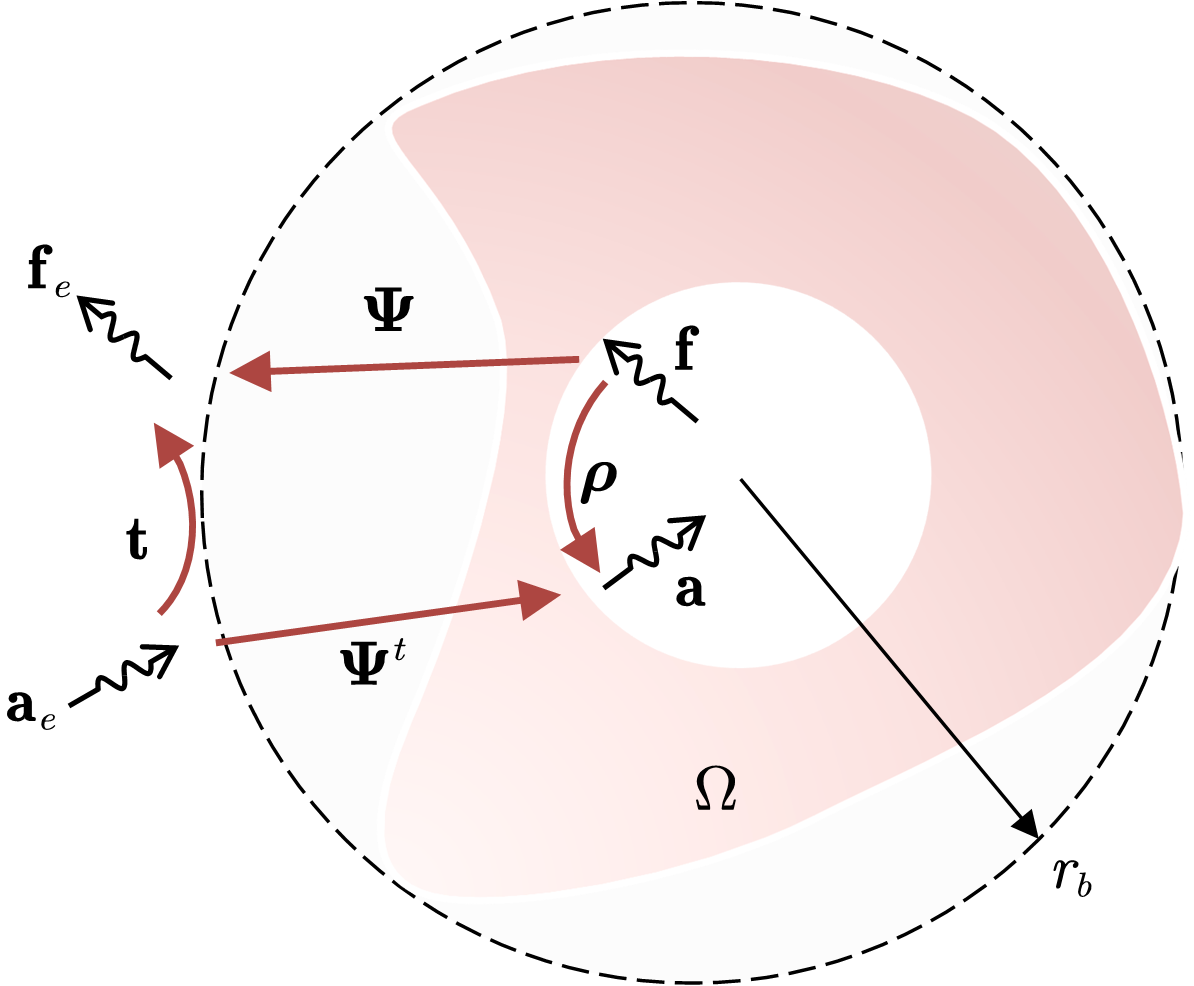}
  \caption{T-matrix representation of electromagnetic waves inside and outside the structure $\Omega$, establishing the bidirectional linear mapping between them.}
  \label{f_Bilinear}
\end{figure}

Another potentially advantageous extension for solving implantable antenna or radome problems is to represent the structure $\Omega$ itself by a generalized transition matrix (T-matrix) in terms of vector spherical waves, as shown in Fig.~\ref{f_Bilinear}. This T-matrix establishes a bidirectional linear mapping between the electromagnetic waves inside and outside $\Omega$:
\begin{equation}
  \label{eq40}
   \begin{bmatrix}
	\mathbf{f}_e\\
	\mathbf{a}\\
\end{bmatrix} =\ \begin{bmatrix}
	\mathbf{t}&		\mathbf{\Psi }\\
	\mathbf{\Psi }^t&		\boldsymbol{\rho }\\
\end{bmatrix} \begin{bmatrix}
	\mathbf{a}_e\\
	\mathbf{f}\\
\end{bmatrix}
\end{equation}

From the second row of \eqref{eq40}, $\mathbf{a}=\mathbf{\Psi}^t \mathbf{a}_e + \boldsymbol{\rho}\mathbf{f}$, which substituted into \eqref{eq21} yields
\begin{equation}
  \label{eq41}
  \begin{cases}	\mathbf{\Gamma v}+\frac{1}{2}\mathbf{R}\left( \mathbf{\Psi }^t\mathbf{a}_e+\boldsymbol{\rho }\mathbf{f} \right) =\mathbf{w}\\	\mathbf{Tv}+\frac{1}{2}\left( \mathbf{S}-\mathbf{1} \right) \left( \mathbf{\Psi }^t\mathbf{a}_e+\boldsymbol{\rho }\mathbf{f} \right) =\mathbf{f}.
    \\\end{cases}
\end{equation}
From the second equation,
\begin{equation}
  \mathbf{f}=\mathbf{M}^{-1}\left[ \mathbf{Tv}+\frac{1}{2}\left( \mathbf{S}-\mathbf{1} \right) \mathbf{\Psi }^t\mathbf{a}_e \right]
\end{equation}
where $\mathbf{M} = \mathbf{1} - \tfrac{1}{2}\big(\mathbf{S}-\mathbf{1}\big)\boldsymbol{\rho}$. The first row of \eqref{eq40} further gives $\mathbf{f}_e = \mathbf{t}\mathbf{a}_e + \mathbf{\Psi}\mathbf{f}$, leading to
\begin{equation}
  \label{eq43}
  \mathbf{f}_e=\mathbf{\Psi M}^{-1}\mathbf{Tv}+\left[ \mathbf{t}+\mathbf{\Psi M}^{-1}\frac{1}{2}\left( \mathbf{S}-\mathbf{1} \right) \mathbf{\Psi }^t \right] \mathbf{a}_e.
\end{equation}

On the other hand, substituting $\mathbf{a}$ and $\mathbf{f}$ into the first equation of \eqref{eq41} yields
\begin{equation}
  \label{eq44}
  \begin{split}
    \mathbf{w}=&\left[ \mathbf{\Gamma }+\frac{1}{2}\mathbf{R}\boldsymbol{\rho }\mathbf{M}^{-1}\mathbf{T} \right] \mathbf{v}\\
    &\qquad+\frac{1}{2}\mathbf{R}\left[ \mathbf{\Psi }^t+\boldsymbol{\rho }\mathbf{M}^{-1}\frac{1}{2}\left( \mathbf{S}-\mathbf{1} \right) \mathbf{\Psi }^t \right] \mathbf{a}_e.
  \end{split}
\end{equation}

Equations \eqref{eq43} and \eqref{eq44} together yield a complete representation of the problem:
\begin{equation}
    \begin{bmatrix}
	\mathbf{w}\\
	\mathbf{f}_e\\
\end{bmatrix}= \begin{bmatrix}
	\tilde{\mathbf{\Gamma}}&		\frac{1}{2}\tilde{\mathbf{R}}\\
	\tilde{\mathbf{T}}&		\frac{1}{2}\left( \tilde{\mathbf{S}}-\mathbf{1} \right)\\
\end{bmatrix} \begin{bmatrix}
	\mathbf{v}\\
	\mathbf{a}_e\\
\end{bmatrix}
\end{equation}
where
\begin{equation}
  \begin{split}
    &\tilde{\mathbf{\Gamma}}=\mathbf{\Gamma }+\frac{1}{2}\mathbf{R}\boldsymbol{\rho }\mathbf{M}^{-1}\mathbf{T}\\
    &\tilde{\mathbf{R}}=\mathbf{R}\left[ \mathbf{\Psi }^t+\boldsymbol{\rho }\mathbf{M}^{-1}\frac{1}{2}\left( \mathbf{S}-\mathbf{1} \right) \mathbf{\Psi }^t \right] \\
    &\tilde{\mathbf{T}}=\mathbf{\Psi M}^{-1}\mathbf{T}\\
    &\tilde{\mathbf{S}}=\mathbf{1}+2\mathbf{t}+\mathbf{\Psi M}^{-1}\left( \mathbf{S}-\mathbf{1} \right) \mathbf{\Psi }^t.
  \end{split}
\end{equation}
This constitutes an alternative hybrid framework, hereafter referred to as the \textit{GSM and T-matrix hybrid method}, since it employs the GSM of the antenna together with the T-matrix of the structure $\Omega$.

A critical step is the determination of the T-matrix subblocks $\mathbf{t}, \mathbf{\Psi}, \boldsymbol{\rho}$, which are obtained from the free-space scattering problem of $\Omega$ alone (\ie without the antenna). Using MoM, these subblocks can be computed as \cite[Appendix C]{ref_hybrid}:
\begin{equation}
  \label{eq47}
  \begin{bmatrix}
	\mathbf{t}&		\mathbf{\Psi }\\
	\mathbf{\Psi }^t&		\boldsymbol{\rho }\\
\end{bmatrix} = \begin{bmatrix}
	-\mathbf{U}_1\mathbf{Z}^{-1}\mathbf{U}_{1}^{t}&		-\mathbf{U}_1\mathbf{Z}^{-1}\mathbf{U}_{4}^{t}+\mathbf{1}\\
	-\mathbf{U}_4\mathbf{Z}^{-1}\mathbf{U}_{1}^{t}+\mathbf{1}&		-\mathbf{U}_4\mathbf{Z}^{-1}\mathbf{U}_{4}^{t}\\
\end{bmatrix}.
\end{equation}

Here, the truncation degree of the spherical wave expansion scales with $k_0 r_b$, so the number of required vector spherical waves grows as $\left(k_0 r_b\right)^2$. For geometrically simple structures, this complexity matches that of the RWG basis functions. However, when $\Omega$ exhibits fine geometric details or high curvature without an increase in overall size, the number of RWG basis functions grows significantly, whereas the number of spherical waves remains unchanged. In such cases, storing $\mathbf{t}, \mathbf{\Psi}, \boldsymbol{\rho}$ is more advantageous than storing $\mathbf{Z}^{-1}$.

We do not claim that the GSM and T-matrix hybrid is universally superior. As the scale of $\Omega$ increases, the evaluation of the subblocks in \eqref{eq47} must also be accelerated using matrix-free methods, which leads to a multiple right-hand-side problem. Its complexity is proportional to the product of the spherical wave count $\left(k_0 r_b\right)^2$ and $N_{\mathrm{it}}\left(k_0 r_b\right)^2\log\left(k_0 r_b\right)$, \ie $N_{\mathrm{it}}\left(k_0 r_b\right)^4\log\left(k_0 r_b\right)$. By contrast, we have already shown that the MoM and GSM hybrid framework can be accelerated to $N_{\mathrm{it}}\left(k_0 r_b\right)^2\log\left(k_0 r_b\right)$ (see Sec.~\ref{Sec_IIIA}). Clearly, for electrically large problems, the MoM + GSM method is more efficient than the GSM + T hybrid method.

A notable exception occurs when $\Omega$ is a sphere (or, at most, an axially inhomogeneous material). In this case, $\mathbf{t}, \mathbf{\Psi}, \boldsymbol{\rho}$ are diagonal and can be obtained analytically \cite{ref_myImplant}, reducing the computational complexity of the GSM + T hybrid method to linear order in $k_0 r_b$. Therefore, the GSM + T hybrid method provides clear advantages in certain special scenarios, whereas for general electrically large problems, the MoM + GSM hybrid remains the more practical and efficient choice.

\section{Numerical Examples}
\label{Sec_IV}

\subsection{Example 1---Spherical Implantable Antenna}

\begin{figure}[!t]
  \centering
  \includegraphics[]{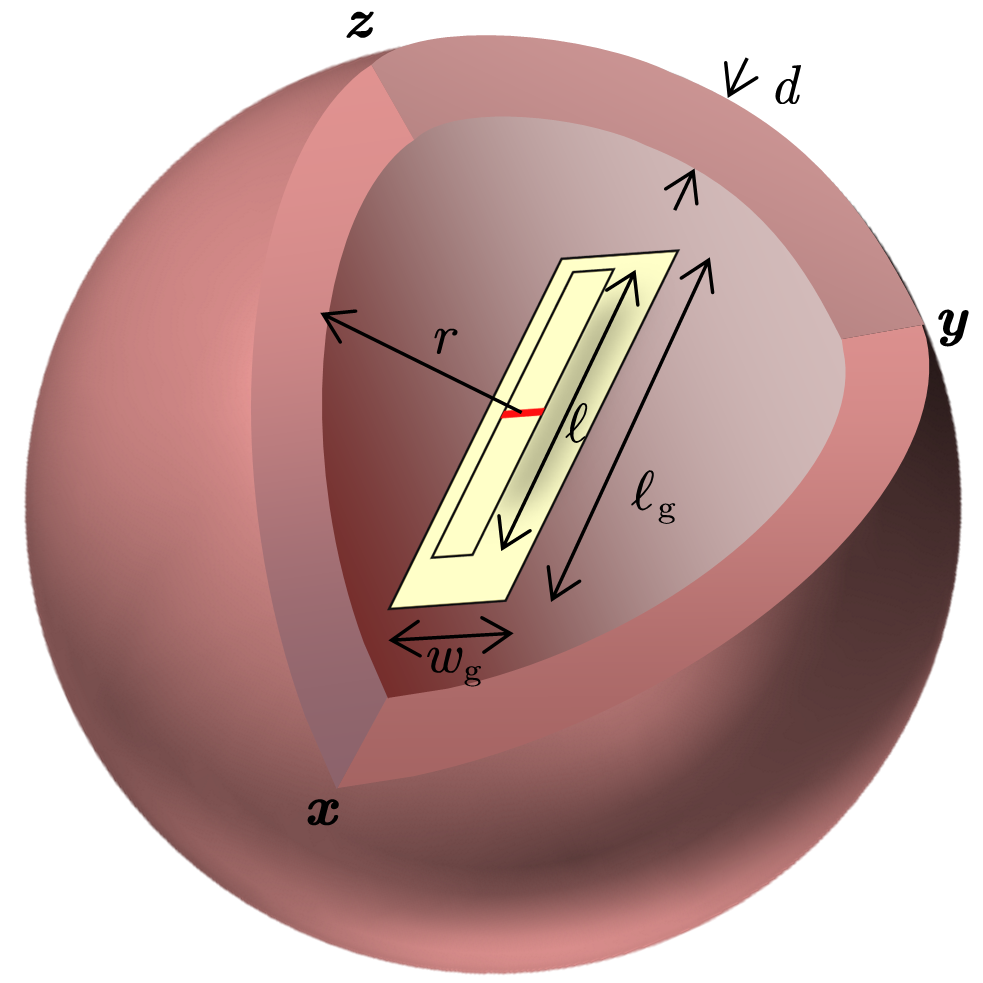}
  \caption{Geometry of the two-element Yagi antenna implanted inside a spherical shell. Dimensions are specified as $w_g=0.25\ell$, $\ell_g=1.25\ell$, $r=0.8\ell$, and $d=0.2\ell$, with the dipole element of width $\ell/10$ placed at a distance $h=0.31\ell$ above the reflecting ground. The dipole is center-fed by a 50~$\Omega$ coaxial line. The spherical shell has a relative permittivity of 5 and a loss tangent of 0.09.}
  \label{f_dipole_sph}
\end{figure}

This example considers the implantable antenna scenario illustrated in Fig.~\ref{f_dipole_sph}, where a center-fed two-element Yagi dipole is embedded inside a lossy spherical cavity. The spherical layer has a relative permittivity of $\epsilon_r = 5$ and a loss tangent of $\tan \delta = 0.09$. In this case, a total of 23,106 RWG basis functions are used to discretize the induced currents on the spherical layer, while 2,014 RWG basis functions are employed to discretize the surface currents on the antenna. With $\iota = 2$ in \eqref{eq12}, the truncation degree is set as $L_{\max}=11$, resulting in 286 vector spherical wave functions being used to expand the antenna fields.

\begin{figure}[!t]
  \centering
  \includegraphics[]{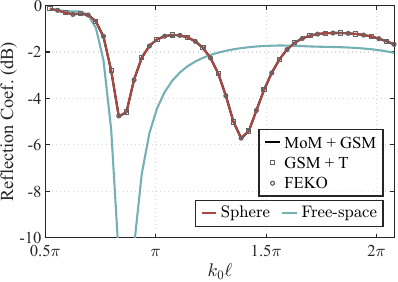} 
  \caption{Reflection coefficient of the antenna in the configuration of Fig.~\ref{f_dipole_sph}. Two resonance peaks are observed at $k_{0}\ell=0.83\pi$ and $k_{0}\ell=1.38\pi$. Results obtained from three different numerical methods are compared.}
  \label{f_Spara_vs_dipoleSphere}
\end{figure}

Figure~\ref{f_Spara_vs_dipoleSphere} shows the port reflection coefficient of the antenna. The curve labeled ``MoM + GSM'' corresponds to the proposed hybrid method, while the points labeled ``FEKO'' are obtained from the commercial full-wave solver FEKO. The two results agree exactly. Compared with the free-space case, the presence of the spherical structure introduces two resonances within the range $k_0\ell \in [0.5\pi, 2\pi]$. This occurs because the dielectric environment shifts higher-order resonances of the antenna toward lower frequencies, consistent with engineering intuition.

\begin{figure}[!t]
  \centering
  \includegraphics[]{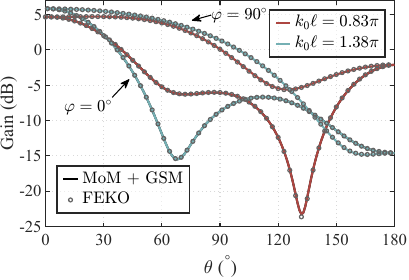} 
  \caption{Gain patterns of the antenna in Fig.~\ref{f_dipole_sph} at its two resonance points.}
  \label{f_Gain_dipoleSphere}
\end{figure}

\begin{figure}[!t]
  \centering
  \includegraphics[]{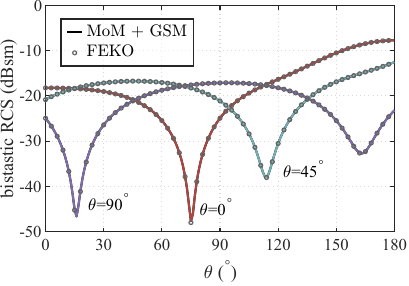} 
  \caption{Bistatic RCS of the antenna in the configuration of Fig.~\ref{f_dipole_sph}, under plane-wave incidence from $\theta=0^{\circ}$, $45^{\circ}$, and $90^{\circ}$.}
  \label{f_RCS_dipoleSphere}
\end{figure}

Figure~\ref{f_Gain_dipoleSphere} illustrates the radiation patterns in the $xoz$ plane ($\varphi = 0^\circ$) and $yoz$ plane ($\varphi = 90^\circ$), while Fig.~\ref{f_RCS_dipoleSphere} shows the bistatic RCS under plane-wave excitations from three different directions. All results match those from FEKO perfectly, further verifying the accuracy of the proposed method.

For this example, evaluating \eqref{eq30} with the MoM and GSM hybrid method required a total runtime of 165.98~s. Specifically, assembling the impedance matrix $\mathbf{Z}$ took 44.27~s, computing its LU decomposition required 116.44~s, evaluating $\mathbf{U}_4\mathbf{Z}^{-1}\mathbf{U}_4^t$ consumed 3.41~s, and obtaining the antenna scattering matrix $\mathbf{S}$ together with inverting $\mathbf{M}$ required only 1.04~s. By contrast, using a purely MoM-based solver such as FEKO, assembling the system matrix took 51.65~s and its LU decomposition required 128.46~s, resulting in a total runtime of 180.11~s.

For a single evaluation, the hybrid method therefore exhibits comparable computational cost to FEKO. However, in parametric studies of the antenna geometry, the precomputed quantities $\mathbf{Z}^{-1}$ and $\mathbf{U}_4\mathbf{Z}^{-1}\mathbf{U}_4^t$ remain unchanged. In such cases, only $\mathbf{S}$ and $\mathbf{M}^{-1}$ need to be re-evaluated, reducing the total runtime to less than 1.5 s per evaluation---over two orders of magnitude faster. This highlights the suitability of the proposed method for performance optimization of implantable antennas.

The GSM and T-matrix hybrid method can also be applied to this scenario, as indicated by the curve labeled ``GSM + T'' in Fig.~\ref{f_Gain_dipoleSphere}. The results are identical to those of the MoM + GSM hybrid method and FEKO. Moreover, since the T-matrix of the spherical structure can be obtained analytically in this case \cite{ref_myImplant}, the total runtime is further reduced to only 1.16 s.

\subsection{Example 2---Conical Radome}

\begin{figure}[!t]
  \centering
  \includegraphics[]{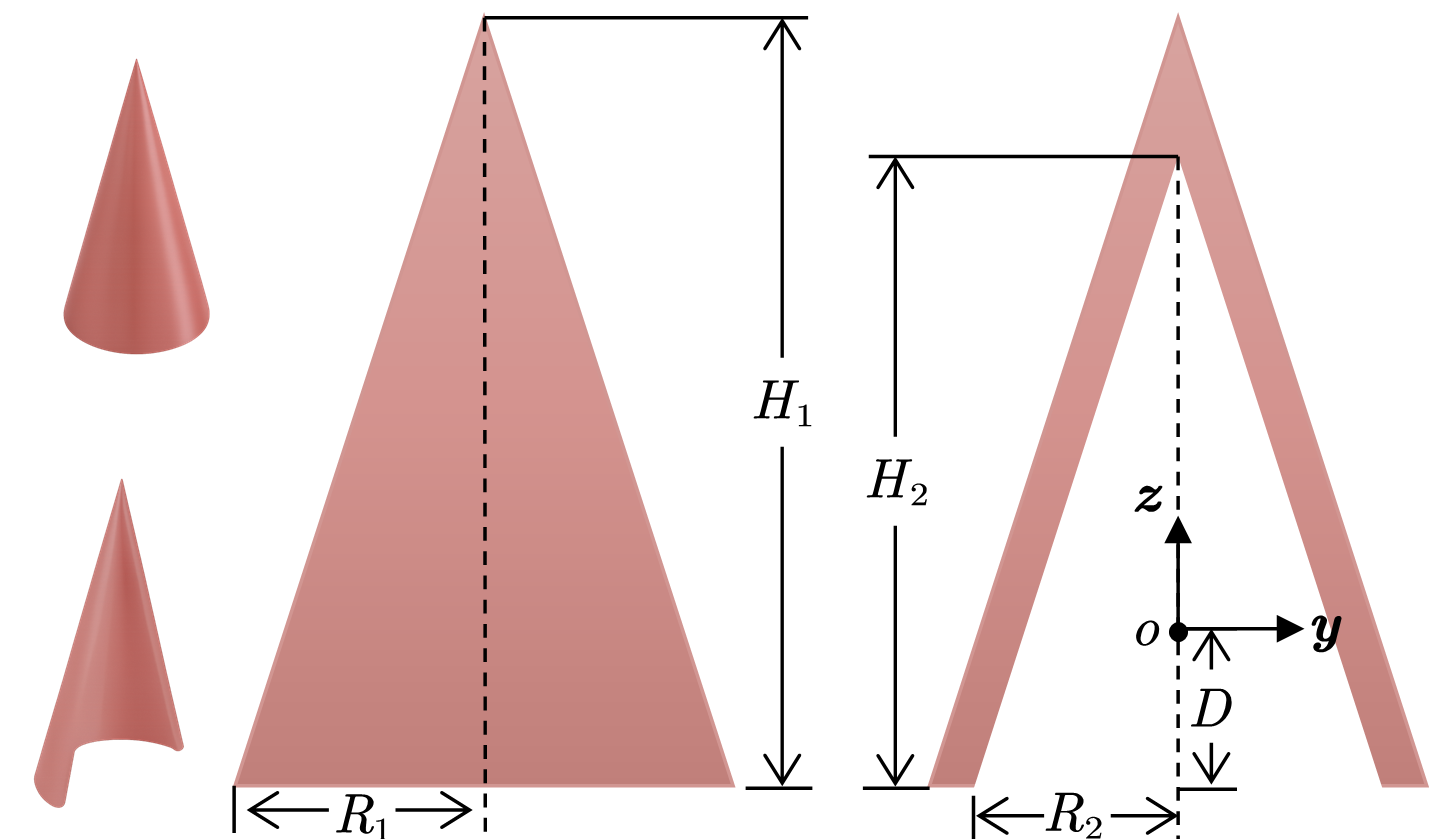}
  \caption{Model of the conical radome. Dimensions are $R_{1}=280$~mm, $H_{1}=630$~mm; $R_{2}=250$~mm, $H_{2}=600$~mm. The coordinate origin is located at point $o$, at a height $D$ above the base of the cone.}
  \label{f_conicalradome}
\end{figure}

The second example continues with the antenna structure used in Fig.~\ref{f_dipole_sph}, with its basic dimensions specified as $\ell=26$~mm, corresponding to resonance near 5~GHz. In this case, a $15 \times 1$ linear array is formed along the $y$-axis, with an inter-element spacing of $d=20$~mm. The outer structure is changed from the spherical shell to a conical radome, as illustrated in Fig.~\ref{f_conicalradome}, with the array centered at the origin $o$. The GSM of the array is synthesized efficiently from the GSMs of individual elements \cite{ref_mysyn_GSM,ref_Fast_Tmat}, thereby avoiding the need for full-wave simulation of the entire array.

\begin{figure}[!t]
  \centering
  \subfloat[]{\includegraphics[]{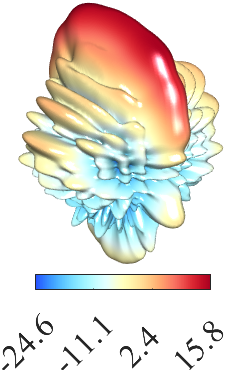}}
  \hfil
  \subfloat[]{\includegraphics[]{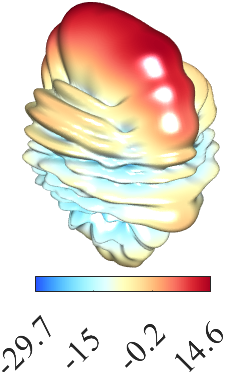}}
  \hfil
  \subfloat[]{\includegraphics[]{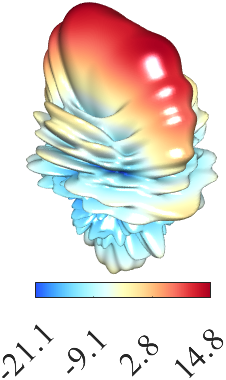}}
  \vfil
  \subfloat[]{\includegraphics[]{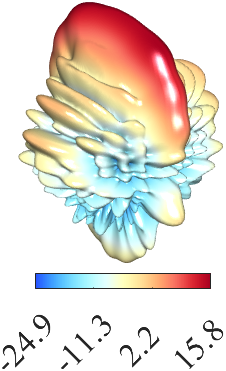}}
  \hfil
  \subfloat[]{\includegraphics[]{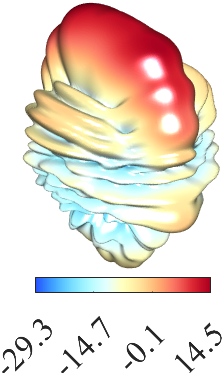}}
  \hfil
  \subfloat[]{\includegraphics[]{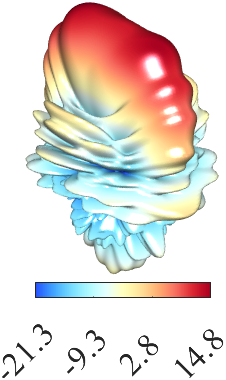}}
  \caption{Radiation gain patterns of the antenna array placed inside the radome. Panels (a), (b), (d), and (e) correspond to $D=210$~mm, where (a) and (d) use uniform in-phase excitation, and (b) and (e) use Taylor distribution excitation. Panels (d) and (f) correspond to $D=120$~mm with Taylor distribution excitation. The top panels show results obtained with the MoM + GSM hybrid method, while the bottom panels show results from FEKO simulations.}
\label{f_Gain_radome}
\end{figure}

We evaluate the array performance when it is placed at a depth of $D=210$~mm inside the radome. Figure~\ref{f_Gain_radome} shows the gain patterns under uniform excitation and Taylor distribution excitation. The top panels present the results of the proposed hybrid method, while the bottom panels provide those obtained with FEKO. The two sets of results exhibit excellent agreement.

In many practical applications, it is often necessary to explore different antenna installation depths and orientations. Within the proposed hybrid framework, such scenarios can be analyzed efficiently because the GSM of the transformed antenna can be obtained from the original GSM using Wigner translation and rotation matrices, without rerunning full-wave simulations. The transformed GSM is given by
\begin{equation}
  \hat{\mathbf{S}}^{\prime}= \begin{bmatrix}
	\mathbf{1}&		\\
	&		\mathcalbf{T} ^t\\
\end{bmatrix} \begin{bmatrix}
	\mathbf{\Gamma }&		\mathbf{R}\\
	\mathbf{T}&		\mathbf{S}\\
\end{bmatrix} \begin{bmatrix}
	\mathbf{1}&		\\
	&		\mathcalbf{T}\\
\end{bmatrix}
\end{equation}
where the transformation matrix $\mathcalbf{T}$ is expressed as the product of the Wigner translation and rotation matrices:
\begin{equation}
  \mathcalbf{T} =\mathcalbf{R} \left( \delta \right) \cdot \mathcalbf{D} \left( \alpha ,\beta ,\gamma \right) 
\end{equation}
with $\delta$ denoting the displacement along the $z$-axis, and $\alpha,\beta,\gamma$ representing the Euler rotation angles. These sparse matrices are available in closed form \cite[Appendix B and Appendix C]{ref_mysyn_GSM}.

\begin{figure}[!t]
  \centering
  \subfloat[]{\includegraphics[]{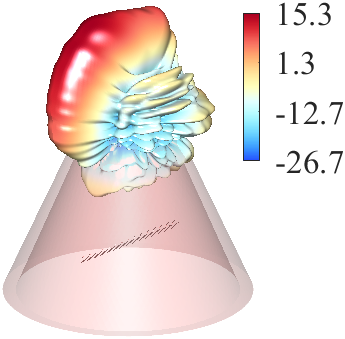}}
  \hfil
  \subfloat[]{\includegraphics[]{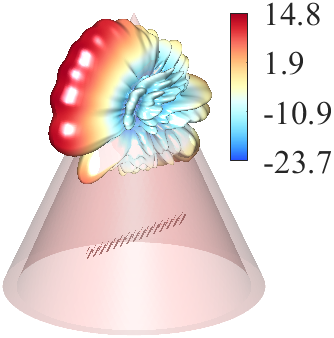}}
  \hfil
  \subfloat[]{\includegraphics[]{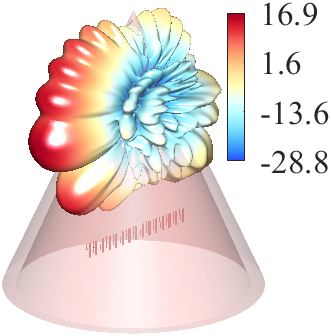}}
  \vfil
  \subfloat[]{\includegraphics[]{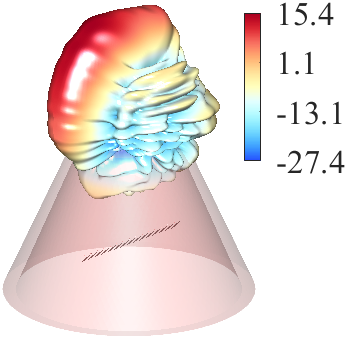}}
  \hfil
  \subfloat[]{\includegraphics[]{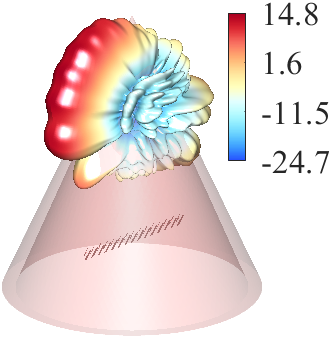}}
  \hfil
  \subfloat[]{\includegraphics[]{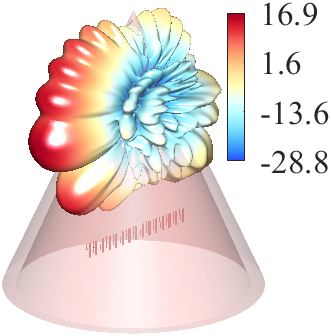}}
  \caption{Radiation gain patterns of the antenna array rotated about the $y$-axis by different angles $\beta$. Panels (a) and (d) correspond to $\beta=30^{\circ}$, panels (b) and (e) to $\beta=60^{\circ}$, and panels (c) and (f) to $\beta=90^{\circ}$. The top panels show results from the MoM + GSM hybrid method, while the bottom panels show results from full-wave FEKO simulations.}
\label{f_Gain_radome_rot}
\end{figure}

To demonstrate the efficiency of this approach, we shift the antenna array along the $z$-axis by $\delta=-90$~mm, corresponding to $D=120$~mm. Under Taylor excitation, Figs.~\ref{f_Gain_radome}(c) and \ref{f_Gain_radome}(f) compare the radiation patterns computed by the hybrid method and FEKO, showing excellent consistency. Building on this, Fig.~\ref{f_Gain_radome_rot} illustrates how the radiation pattern changes as the antenna array is tilted by different angles $\beta$ around the $y$-axis. Again, the agreement with FEKO results validates the accuracy of the hybrid method.

It is worth emphasizing that in FEKO simulations, every positional or orientational adjustment of the array requires a complete re-simulation, which is computationally expensive. By contrast, the proposed hybrid framework provides significant efficiency advantages in addressing such scenarios.

\subsection{Example 3---Reflector Antenna}

\begin{figure}[!t]
  \centering
  \includegraphics[]{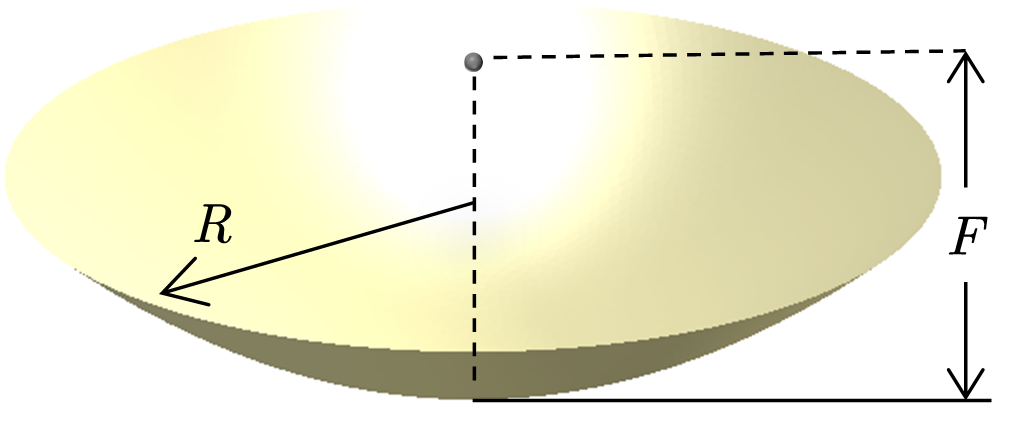}
  \caption{Parabolic reflector model with focal point at $F=180$~mm outside the reflector and an aperture radius of $R=300$~mm.}
  \label{f_parabolid}
\end{figure}

\begin{figure}[!t]
  \centering
  \subfloat[]{\includegraphics[]{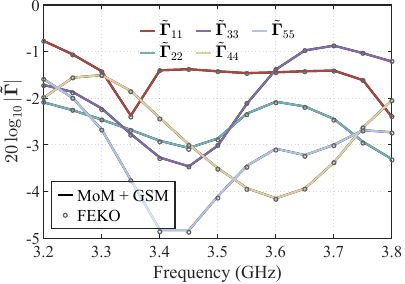}}
  \vfil
  \subfloat[]{\includegraphics[]{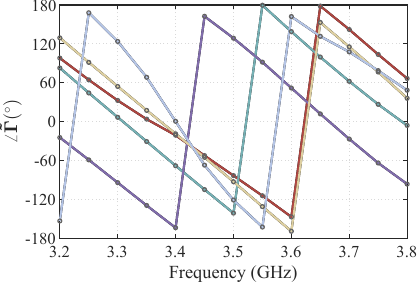}}
  \caption{Reflection coefficients of the horn antenna for five operating modes: (a) magnitude (dB) and (b) phase.}
\label{f_SparadB_Horn_Refl}
\end{figure}

The final example investigates a parabolic reflector, as illustrated in Fig.~\ref{f_parabolid}. A horn feed, identical in dimension to that used in \cite{ref_myGSM}, is placed at the focal point to illuminate the reflector. Within the frequency band of 3.2-3.8 GHz, the horn supports five operating modes. The MoM and GSM hybrid method is employed to compute the port $S$-parameters in terms of both magnitude and phase, as shown in Figs.~\ref{f_SparadB_Horn_Refl}(a) and \ref{f_SparadB_Horn_Refl}(b). The results are in exact agreement with those obtained from full-wave FEKO simulations, thereby demonstrating the applicability of the proposed method to reflector antennas.

\begin{figure}[!t]
  \centering
  \includegraphics[]{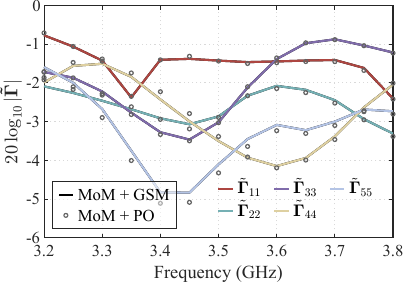} 
  \caption{Same configuration as in Fig.~\ref{f_SparadB_Horn_Refl}, but with the parabolic reflector approximated using the full-coupled PO formulation of Sec.~\ref{Sec_IIIB}. The plots show the magnitudes of the $S$-parameters for the five modes.}
  \label{f_SparadB_Horn_Refl_PO}
\end{figure}

The reflector surface can also be modelled using the PO method, as discussed in Sec.~\ref{Sec_IIIB}. This circumvents most of the computational and storage cost associated with matrix elements involving the reflector, which is particularly critical for the performance evaluation of large reflector antennas. In this example, applying the full-coupled PO formulation derived in Sec.~\ref{Sec_IIIB} yields the curves shown in Fig.~\ref{f_SparadB_Horn_Refl_PO}. A comparison with the MoM + GSM hybrid framework reveals a certain level of deviation, attributable to the high-frequency approximation inherent in the PO method. Overall, however, the derived PO-based algorithm demonstrates acceptable accuracy in this case.

\section{Conclusion}

In this work, we have developed a hybrid framework combining the method of moments (MoM) and the generalized scattering matrix (GSM) for the efficient and accurate analysis of antennas interacting with large surrounding structures. By representing the electrically detailed antenna through a compact GSM description while solving the electrically large but smoother environment with MoM, the framework achieves a clean separation of scales and preserves full electromagnetic coupling. This formulation enables remarkable computational savings by allowing antenna or environment representations to be precomputed and reused across multiple scenarios.

A central contribution of this work lies in its versatility: the same foundation seamlessly evolves into GSM-PO and GSM + T-matrix hybrids, each tailored to specific applications such as reflector antennas or canonical scattering bodies. The proposed approach thus constitutes not merely a single algorithm, but a unified paradigm for multiscale antenna modeling that bridges fine structural details with large-scale environments.

Extensive validations—including implantable antennas, radome-protected arrays, and reflector systems—demonstrate excellent agreement with full-wave solvers while reducing computational costs by orders of magnitude in design and optimization tasks. These results confirm that the hybrid MoM-GSM framework provides both the accuracy demanded by full-wave methods and the efficiency required for practical engineering. Looking forward, its modularity and compatibility with recent advances in GSM theory open a promising pathway toward scalable modeling of increasingly complex antenna systems in diverse real-world environments.

\end{document}